\documentclass[prl,showpacs,showkeys,twocolumn,10pt]{revtex4-1}
\usepackage{amsfonts,bbold,amsmath,amssymb,graphicx,epstopdf,verbatim,dsfont,color}
\usepackage[english]{babel}

\begin{document}

\title{Quantum equivalence, the second law and emergent gravity}
\author{Dries Sels}
\email{dries.sels@uantwerpen.be}
\affiliation{TQC, Universiteit Antwerpen, Universiteitsplein 1, B-2610 Antwerpen, Belgium}
\author{Michiel Wouters}
\affiliation{TQC, Universiteit Antwerpen, Universiteitsplein 1, B-2610 Antwerpen, Belgium}
\date{\today}

\begin{abstract}
Since the advent of quantum mechanics we have mainly been concerned with its predictions from the perspective of an external observer. This is in strong contrast to the theory of general relativity, where the physics is governed by the intrinsic properties of space-time. At the same time, the precise relation between space-time and quantum mechanics is still one of the greatest problems of theoretical physics. This immediately raises the question on the completeness of our understanding of quantum mechanics. Here we address the problem by making an intrinsic analysis of observables in generic quantum systems. We show that there is an extreme fine tuning problem for the emergence of physics from the Hilbert space dynamics. However, for any initial condition and Hamiltonian, there exists a special set of observables. We show that these observables are intimately linked to the natural configuration space in which an area law for the entanglement is inevitable. We argue that this implies emergent gravity.     
\end{abstract}
\maketitle

At the heart of our understanding of space-time, the theory of general relativity, lies the equivalence principle and with it the prominent role played by the observer. The equivalence principle highlights how seemingly different situations for an external observer are fully equivalent for an internal observer, a feature that is naturally embodied in the geometry of space-time. Likewise, the observer has played a prominent role in quantum mechanics from the start, e.g. in Schr\"odinger's cat experiment. Unfortunately, the special role of the observer has troubled, rather than strengthened, the theory and it has led to many different interpretations and solutions to the quantum measurement problem \cite{bell}. The lack of consensus on the latter \cite{quant_int} is a clear sign of missing insight in the emergence of physics from quantum evolution in Hilbert space. Here, we will address this problem by making an intrinsic analysis of the observables in a generic quantum evolution.

A prominent role in our study will be played by the entropy, a quantity that is central \cite{bekenstein}, but not fully understood, in quantum gravity.
While entropy production in open quantum systems is well understood~\cite{Tasaki,GoldsteinTypicality,Popescu,Linden,ReimannOpen} on the basis of the von Neumann entropy $S= {\rm tr} (\rho \ln \rho)$, it remains constant under unitary time evolution in closed systems. For a fundamental law, that should be valid for the universe as a whole, this is not very satisfactory. The second law only appears naturally when the initial state is taken to be separable in configuration space and when the entropy is computed from the local reduced density matrices. This indicates a deep connection between space, time (entropy production) and a specific initial condition. Using recent results on quantum quenches \cite{polk_rmp,eisert_rmp}, we will show that this scenario is  universal: for every macroscopic quantum evolution, a natural configuration space exists such that the initial state is separable. Moreover, the time evolution can always be described by a Hamiltonian with local couplings, for which an area law scaling of the entanglement entropy holds for short evolution times. We will argue that this mathematical result implies emergent gravity.

\section{Observation and time}

In order to understand the physics that can emerge from quantum mechanics, it is important to be clear on the role played by the trinity of Hamiltonian, initial condition and observables. It is quite remarkable that quantum mechanics is invariant under unitary transformations and that the only time evolution in the theory is a unitary transformation of the wave function. This implies that it should be fundamental impossible to have access to all observables. The converse immediately leads to the conclusion that time does not exist or that time travel is possible. In order to travel backwards for a time $\tau$, we would just have to measure the observables $\exp(-iH\tau) \hat O \exp(iH\tau)$ instead of the observable $\hat O$. 

One could also say that time is meaningless, unless one has specified the observables of interest. Indeed, if only the Hamiltonian is specified and no other observables, the phases of the energy eigenstates have no meaning. The Hamiltonian can be written as 
\begin{equation}
H=\sum_n \epsilon_n | n \rangle \langle n |, 
\label{eq:Hamiltonian}
\end{equation}
which is invariant under the unitary transformation $| n \rangle \rightarrow e^{i\theta_n} | n \rangle$. But this transformation is exactly of the same form as the time evolution. One way to break this symmetry is by specifying an observable that does not commute with the Hamiltonian.

Any discussion of time evolution in quantum mechanics should therefore be based on a specific, restricted, set of observables. In an experimental context, the relevant observables are dictated by experimental limitations.
These are formalised by quantum field theory, where the Hamiltonian and observables are intimately related operators, since both have simple algebraic expressions in terms of the field annihilation and creation operators. 
From a fundamental point of view, this tacit assumption of a simple relation between observables and kinematics however raises the question on the completeness of our understanding. 

\section{Fine tuning}
Some recent works on the thermalisation in closed quantum systems have come to remarkable conclusions in this respect. It was shown that that very mild restrictions on the precision of measurements are sufficient to prove that all observables for an overwhelmingly large fraction of time take their thermal equilibrium values \cite{ReimannRealistic,ReimannKastner,Short,ShortFinite}. It was also proven \cite{Malabarba, GoldsteinTime1, GoldsteinTime2} that for typical observables, the time scale for which large nonequilibrium fluctuations persist is extremely short, of the order of the Boltzmann time $\tau_{B} = \hbar/k_B T$, where $T$ is the equilibrium temperature. 

\begin{figure}[tbp]
\centering
\includegraphics[width=0.5\textwidth]{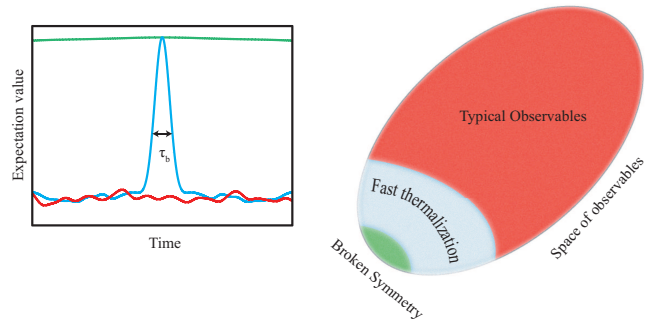}
\caption{The space of all observables can be divided in (red region) the typical observables that are thermal on the considered time interval; (blue region) the fast thermalising observables and (green region) slowly decaying observables, related to broken symmetry. An example of corresponding expectation values are depicted in the same color in the left panel.}
\label{fig:spaceobserve}
\end{figure}

\begin{figure}[tbp]
\centering
\includegraphics[width=0.4\textwidth]{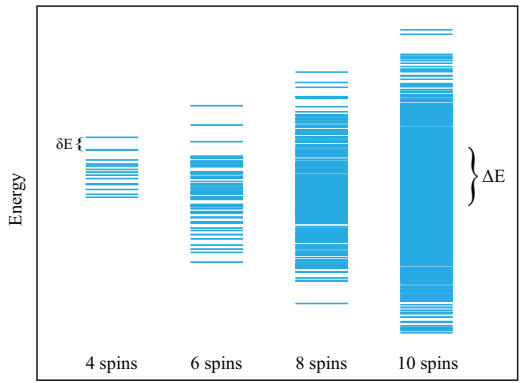}
\caption{Scaling of the many-body spectrum of a typical Hamiltonian with system size. The spectrum becomes exponentially dense with a linear growth of the system size.}
\label{fig:spectrum}
\end{figure}

Those results are readily understood by inspection of the expectation value of an operator $\hat O$ for a pure state, that evolves under the Hamiltonian \eqref{eq:Hamiltonian} 
\begin{equation}
\left\langle  O(t) \right\rangle =\sum_{n,m} c_n O_{nm} c_m  e^{i(\epsilon_n-\epsilon_m)t+i (\phi_m - \phi_n)},
\label{eq:expectO}
\end{equation}
where $O_{nm}=\left\langle n| O| m \right\rangle$ and the initial phases are denoted by $\phi_n$. 
 A typical observable is thermalised because there is no specific relation between the initial phases $\phi_{n}$. Summing over all random phases gives a vanishing expectation value for those observables. Even if the initial phases are chosen appropriate, the sum \eqref{eq:expectO} will in general decay quickly as a function of time, because every component in the sum oscillates at a different frequency $\omega=\epsilon_n-\epsilon_m$, which causes rapid dephasing on the Boltzmann time, that is inversely proportional to the energy width of the state $\tau_B = 1/\Delta E$, see Fig.~\ref{fig:spaceobserve} and~\ref{fig:spectrum}. 
These works solve the problem of thermalisation in closed quantum systems, but at the same time raise the issue of understanding how a nonequilibrium state can exist for an extended period of time, as it is the case for the observables that we are used to in physics. 

A first important result in this respect was the proof that for any quantum system, there exist observables that take a time that is exponentially large in the system size to thermalise \cite{ShortFinite}. The existence of such an operator can be easily identified from expression \eqref{eq:expectO}. For the observable with matrix elements 
\begin{equation}
O_{nm} = e^{-i(\phi_n-\phi_m)}  \delta_{n+1,m}
+  e^{i(\phi_n-\phi_m)} \delta_{n-1,m}
\label{eq:O_SS}
\end{equation}
which couples neighbouring energy levels, it will take a time $t=1/\delta \epsilon$ for the observable to dephase, that is exponentially long in the system size ($\delta \epsilon$ is the typical energy spacing between neighbouring levels, see Fig.~\ref{fig:spectrum}).
This is again not a property of physical observables, with the important exception of spontaneous symmetry breaking.  
Between these two extremes, observables with intermediate relaxation times exist as well, but a crucial lesson is that the number of observables with slow relaxation times is exponentially smaller than the number of fast relaxing ones, as schematically represented in Fig.~\ref{fig:spaceobserve}.
Moreover, also the slow variables take on their thermal expectation values for almost all times after an initial transient. Therefore the statistical probability to a system out of equilibrium is exponentially small.

All this illustrates that an extreme fine tuning between a quantum state and observable is needed in order to obtain any dynamics.
It is tempting to look at the so-called fine tuning problems in cosmology and elementary particle physics from this perspective. 
When one is puzzled by the small probability for the existence of our universe, one does so from the perspective of the observables that we consider to be physical (built of field operators of stable particles). This explains the fine tuning problems: the initial condition of the universe had to be very special with respect to our electron and proton field operators in order to see the present structure.

The thermalisation results however offer a solution: for any initial condition and Hamiltonian, there exists a set of special observables that relax slowly. This is clear from \eqref{eq:O_SS}: the slow observables depend on the initial condition, but for any initial condition a slow observable can be constructed. We will argue below that those slow observables are intimately connected to the emergence of space. But before proceeding with this task, we will present our argument for the equivalence of Hamiltonian evolutions of macroscopic systems.

\section{Quantum equivalence}
Consider the unitary time evolution of a state
\begin{equation}
| \Psi(t) \rangle = e^{-i H t} |\Psi(0)\rangle =  \sum_n c_n   e^{-i \epsilon_n t} | n \rangle.
\end{equation}
The phases of the initial condition $c_n$ have now been absorbed in the definition of the eigenstates $| n \rangle $: the intrinsic quantum evolution is then clearly independent of the initial condition. It should be understood that this is the case, only because we have not yet specified any other observable of interest.

The dynamics is then determined by the spectrum of the Hamiltonian only. The individual states of this spectrum are not resolved as long as the time is not exponentially long ($t < 1/\delta \epsilon$). The only microscopic element that we remain with is the density of occupied states $\rho(E)$. The following conclusion is thus unavoidable: all quantum systems with the same density of occupied states, on a scale set by the evolution time, are equivalent.

Surprisingly, this equivalence principle leads to the conclusion that a `quantum theory of everything' consists merely of the specification of a density of states. Studies in many body physics have shown that the density of occupied states is generically of Gaussian form \cite{venuti,santos_strength}. It thus appears that all known interacting field theories essentially lead to equivalent dynamics (of course the precise relation between different systems is typically enormously complicated). 
Under the assumption that quantum mechanics is valid for the universe as a whole, Occam's razor principle suggests to conjecture that it has a Gaussian density of occupied states.

We then conclude that on the level of Hilbert space dynamics, a typical quenched many body system is equivalent to the whole universe. Note that in this reasoning even the difference between systems made of fermionic and bosonic particles disappears. The reason why they appear so different to us is that we have a completely different experimental access to them, through fermionic or bosonic field operators respectively. The way one looks at a system thus determines its properties rather than the intrinsic nature of the system itself. 

\section{Natural configuration space and second law}

The quantum equivalence principle offers us the freedom to choose the pair initial state/Hamiltonian that simplifies the identification of the observables that feature a long relaxation time. In order to make contact with a physical configuration space, we wish to write our Hilbert space as a tensor product of subspaces $\mathcal H = \bigotimes_{j} \mathcal H_j$ in such a way that the local operators relax slowly to equilibrium. 

\begin{figure*}[tbp]
\centering
\includegraphics[width=\textwidth]{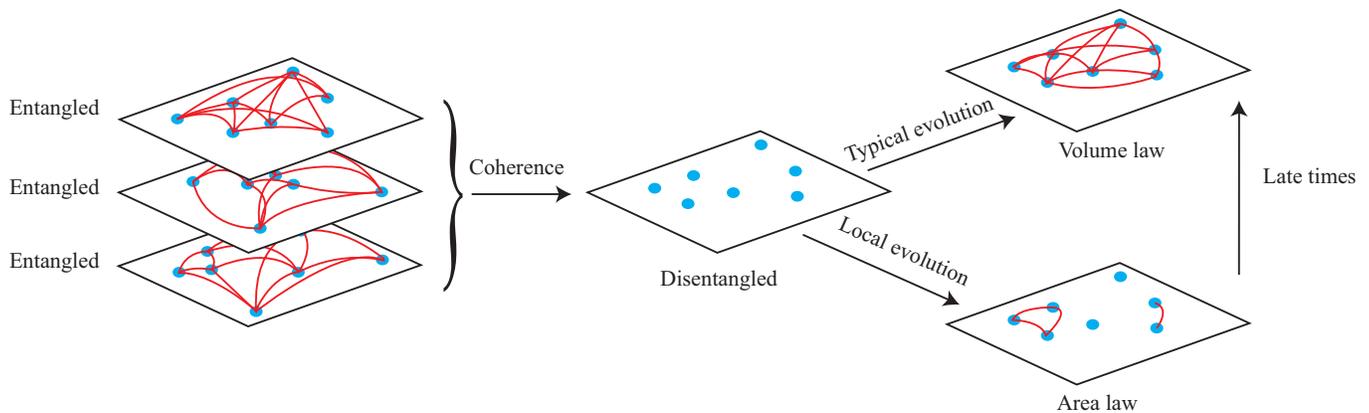}
\caption{While excited eigenstates generically show a volume law for the entanglement scaling, the coherent initial state can be chosen to be disentangled. A typical Hamiltonian evolution entangles all points within the Boltzmann time. Under a local Hamiltonian evolution however, the entanglement scales initially with an area law, indicating gravitational interactions between emergent quasi-particles.}
\label{fig:space}
\end{figure*}

For a generic decomposition of Hilbert space, one finds however that local observables are thermal and that the entanglement entropy is proportional to the volume of the region \cite{hayden, dahlsten}. Note here the correspondence with the fact that typical observables are thermalised. Moreover, for a generic decomposition, the Hamiltonian is highly non-local, such that even when initially the entanglement entropy vanishes, it will become extensive on the order of the Boltzmann time, together with the thermalisation of the local observables (see Fig.~\ref{fig:space}). Again, this is in correspondence with the short thermalisation time for typical observables. Only when the Hamiltonian is local in the configuration space, an initially separable state will lead to a slow growth of the entanglement entropy in time. In this situation, one can show that the entanglement entropy grows typically linear with time and quite generally features an area law \cite{eisert_rmp}
\begin{equation}
S(X) \sim t A(X).
\end{equation}
Only then, the local observables coincide with the small subset of slowly decaying observables.
Note that the second law, interpreted as the increase of entanglement entropy, comes out immediately, thanks to the special relation between initial condition and the natural configuration space.

Given the essential connection between spontaneous symmetry breaking and slow thermalisation, it is better to choose a model where one knows the broken symmetry. In that way, the physical order that will persist for long times is immediately apparent.
One could think of a quantum spin system, starting in a separable state. 
We have thus come to a quite specific model for our universe, but thanks to the quantum equivalence principle, this is not at the expense of generality.

The celebrated Lieb-Robinson bound \cite{liebrobinson} implies a finite propagation speed for the entanglement.
In a cosmological interpretation, the finite speed for the spreading of entanglement implies that a given point in space only has information about a finite part of the universe. This corresponds with an observable universe that is only a part of the whole. Initially, all points in space are identical and they do not share any information. Consequently there is no reasonable definition of space. It has furthermore been shown by Van Raamsdonk \cite{vanraamsdonk} that, in the context of gauge-gravity correspondence \cite{thooft,maldacena}, separable quantum states lead to disconnected space-times.

With our condensed matter construction, we come to a universe that is homogeneous in space, in analogy to Volovik's Helium droplet universe \cite{volovik}. Actually, our only additional ingredient to Volovik's universe is the identification of the initial condition as a separable state. It is this initial condition that allows for the creation of quasi-particles {\em ex nihilo}. As time goes on, entanglement spreads, which can be described as the creation and propagation of quasi-particles.  

\section{Gravity from entanglement scaling}

We highlighted above the connection between a slow relaxation of local observables and an area law for the entanglement entropy. From a physical point of view, this is actually a rather surprising situation, because it does not correspond to the scaling of a collection of independent quasi-particles, that is extensive. We are thus led to the conclusion that the emergent quasi-partices should be correlated. In our previous work, we have shown that correlations due to an initial condition can be described in a statistical generalised Gibbs description \cite{rigol,jaynes} as due to fictitious interactions \cite{gge}. The  area law for the entropy and universality of this interaction (it should act on all types of quasi-particles) leads us to the conjecture that this `spooky' interaction is gravity.

It was actually the suggestive relation between area laws for the ground state of condensed matter systems and black holes \cite{bombelli,srednicki_ar} that formed a major motivation for the study of entanglement entropy. We wish to stress that the area law is here the consequence of the short time evolution (times for which the observable universe is smaller than the whole system) and not of the fact that the system is in the ground state.
In addition, there are thermodynamic indications that gravity is related to a negative contribution to the entropy. It has long been known that gravitational systems have a negative specific heat, both for systems of Newtonian gravitating masses \cite{lynden} and for black holes \cite{bekenstein}. The idea that gravity and more generally Einstein relativity has an entropic origin was already introduced two decades ago by Jacobson \cite{jacobson}. More recently, gravity was argued to be an entropy-related force on the basis of holographic arguments by Verlinde \cite{verlinde}. Inspired by Verlinde's work, there have been several other works that speculate on the connection between entanglement entropy and gravity \cite{lee,mach,lashkari}.

In contrast to the previous works on entropic gravity, our analysis did not require any assumptions in addition to unitary quantum evolution. Verlinde's analogy with colloid and bio-physical systems is rather confusing in this respect, because it suggests the association decoherence effects to gravitational interactions \cite{kobak}. In our view, gravity is rather a consequence of missing entropy than due to an additional entropic process. It is because of the slow growth of entropy according to an area law that the quasi-particles have to be correlated in space, which is perceived as a gravitational interaction.

\section{Conclusions}

Let us recapitulate the main results of our analysis.
We argued that the time evolution in Hilbert space of macroscopic quantum systems is equivalent. The fact that most observables are most of the time equal to their thermal expectation value was interpreted as a fine tuning problem: a precise relation is required between the wave function and an observable in order to show deviations from the ergodic average. 

We used the quantum equivalence principle in order to construct a model where the local observables relax slowly. 
It was argued that the initial state should be separable in the natural configuration space and that the Hamiltonian is local. An area law for the entanglement is then inevitable. The usual second law of thermodynamics is a direct consequence of the separability of the initial state. A phenomenological analyses of the entanglement entropy naturally led us to the identification of classical gravity. 

With this paper, we have only scratched the surface of `quantum quench cosmology' and many questions remain unanswered. For example, we suspect that black holes are related to partial thermalisation. The no-hair theorem for black holes \cite{nohair} could then be a consequence of the eigenstate thermalisation hypothesis \cite{srednicki,deutsch,rigol2} for many body systems.  
In general, we hope that our equivalence principle will lead to new insights in the fundamental structure of nature, based on analogies with condensed matter systems, a strategy that has proven so successful the past \cite{anderson,higgs,brout,ghk}.

\acknowledgements
D.S. acknowledges support of the FWO as post-doctoral fellow of the Research Foundation - Flanders. M. W. acknowledges financial support from the FWO Odysseus program.

\end{document}